\documentclass[4pt]{article}

\usepackage{fullpage}
\usepackage{graphicx}
\usepackage{amsmath}
\usepackage{algpseudocode}
\usepackage{verbatim}

\begin{document}

\title{Design, Implementation and Evaluation of MTBDD based Fuzzy Sets and Binary Fuzzy Relations}

\author{Hamid A. Toussi$^{*}$\\
        Bahram Sadeghi Bigham$^{**}$\\
        *Young Researchers and Elites Club, Mashhad  Branch\\Islamic Azad University, Mashhad, Iran\\hamidt@mshdiau.ac.ir\\
        **Department of Computer Science \& Information Technology\\Institute for Advanced Studies in Basic Sciences(IASBS)\\Zanjan, Iran\\
              b\_sadeghi\_b@iasbs.ac.ir
}

\date{}

\maketitle

\begin{abstract}
For fast and efficient analysis of large sets of fuzzy data, elimination of redundancies in the  memory representation is needed. We used MTBDDs as the underlying data-structure to represent fuzzy sets and binary fuzzy relations. This leads to elimination of redundancies in the representation, less computations, and faster analyses. \\
We have also extended a  BDD package (BuDDy) to support MTBDDs in general and fuzzy sets and relations in particular. Different fuzzy operations such as max, min and max-min composition were implemented based on our representation.\\
Effectiveness of our representation is shown by applying it on fuzzy connectedness and image segmentation problem.\\
Compared to a base implementation, the running time of our MTBDD based implementation was faster (in our test cases) by a factor ranging from 2 to 27. Also, when the MTBDD based data-structure was employed, the memory needed to represent the final results was improved by a factor ranging from 37.9 to 265.5.
\end{abstract}


\section{Introduction}
\label{introduction}
ROBDDs have been used in hardware community for model checking and circuit verification \cite{symbol_mc}. It has a variety of applications. For example, It is used in compiler community for efficient points-to analysis \cite{pt_bdd,cs_bdd,pt_xbdd} and it is also used in image processing \cite{image_bdd,image_bryant}.
%H. Nakahara used different variants of ROBDDs (MTBDDs and MTMDDs) to represent logic circuits efficiently \cite{circuit_mtbdd}. 
%The proposed method was claimed to be more efficient than the original ROBDD representation. 

Efficient representation of fuzzy sets and relations can be of great importance for analysing large sets of fuzzy data.\\In this work, design and implementation of a MTBDD based data-structure for representing fuzzy sets and binary fuzzy relations is investigated. Our main idea is to use MTBDDs \cite{mtbdd} as the underlying data-structure.

MTBDDs has been used to represent arrays and graphs \cite{mtbdd,add}. Clark et al. discussed representation of 2-dimensional arrays and vectors by using MTBDDs. However, they did not provide any implementation. Also they used shadow nodes to simplify their algorithms. Our idea is incorporated into a modern BDD library (BuDDy \cite{buddy}) without shadow nodes. Shadow nodes increase the size of MTBDDs and thus make the implementation less efficient. Instead, level attribute already presented in the BDD library, is used.

R. Iris Bahar used MTBDDs to perform matrix multiplication and also solve all pairs shortest paths problem \cite{add}. D. Bugaychenko used MRBDD for doing probabilistic model checking \cite{model_checking_mrbdd}. Vaclav Dvorak discussed the implementation of MTBDDs on special purpose Decision Diagram Machines (DMMs) \cite{branching_programs}. D. Yu. Bugaychenko and I. P. Soloviev proposed MRBDD (Multi-root decision diagram) data-structure to represent integer functions. In their representation, a finite-valued function is represented by a list of $k$ different ROBDDs (or $k$ roots as they suggested) thus an assignment maps to a binary string of $0$s and $1$s of length $k$ instead of just a $0$ or $1$. The resulted binary string should be decoded to a certain value. This value is the output of the function for the assignment. The list of ROBDDs which constitute the MRBDD share isomorphic sub-graphs (every sub-graph is also a ROBDD) \cite{mrbdd}.

In our implementation, because of the way that BuDDy allocates nodes, no two isomorphic ROBDD is ever allocated twice so our work does not just share sub-graphs among a set of ROBDDs belonging to a single MRBDD but among all ROBDDs in memory. This is also discussed in Section~\ref{buddy_sec}. Generally speaking, sharing is more pervasive in MRBDDs compared to MTBDDs since there are just two terminals instead of a set of terminals. However, implementation of operations on MTBDDs is straightforward because terminals are shown explicitly. This is not the case with MRBDD and for any operation, a correspondence between operation on output values and equivalent operation on binary encoding of the output values must be defined. Summation and multiplication on matrices which  are represented by MRBDDs are explained in \cite{mrbdd}.\\
An early version of our work was published in proceeding of International Conference on Computer, Information Technology and Digital Media \cite{citadim}.

We have evaluated our data-structure by running max-min composition and a routine that solves fuzzy connectedness over different binary fuzzy relations. These relations are obtained from different images. Results are compared with a base implementation which uses 2-dimensional arrays to represent binary fuzzy relations.\\
Considering max-min composition routine, MTBDD based implementation ran 164--1328$\times$ faster depending on number of pixels in input image (i.e. its size), values stored in every pixel and desired precision in the output.
In the other evaluation which solved fuzzy-connectedness problem, MTBDD based implementation was 18 -- 27$\times$ faster when number of distinct membership values that can appear in relations is limited to 11 values. In all cases we have far better memory consumption when MTBDD based implementation is used. See Section~\ref{evaluation} for more detail.

Our major contributions in this work are:
\begin{itemize}
\item Describing representation and manipulation of fuzzy sets and binary fuzzy relations based on MTBDDs
\item Extending BuDDy library to support MTBDDs in general and fuzzy sets and binary fuzzy relations in particular
\item Evaluation of our implementation by solving fuzzy connectedness problem 
\end{itemize}

An introduction to ROBDDs, BuDDy and MTBDDs comes in Section~\ref{background}. The way that BuDDy is extended is discussed in Section~\ref{extend_buddy_sec}. Representation and manipulation of MTBDD based fuzzy sets and binary fuzzy relations are discussed in Sections~\ref{mtbdd_fuzzyset_sec} and~\ref{mtbdd_fuzzyrelation}. The empirical evaluation is given in Section~\ref{evaluation} and finally, in Conclusion, we discuss possible future directions.

\section{Background}
\label{background}
\subsection{Binary Decision Diagrams}
\label{bdd_sec}
\begin{figure} [ht]
\begin{center}
\includegraphics[scale=0.8]{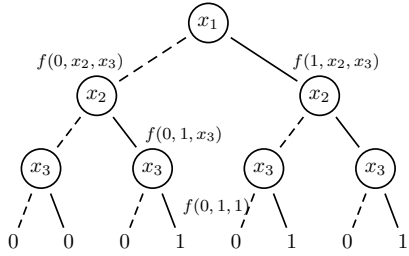}
\end{center}

\caption{
A completely unreduced BDD which represents the boolean function $f= \neg x_1.x_2.x_3 + x_1. \neg x_2.x_3 + x_1.x_2.x_3$.
}
\label{bdd_unreduced}

\end{figure}

BDD is a data structure for representing boolean functions compactly.\\
A completely unreduced BDD is shown in Figure~\ref{bdd_unreduced} which represents the boolean function $f= \neg x_1.x_2.x_3 + x_1. \neg x_2.x_3 + x_1.x_2.x_3$. Behind some of the nodes, their associated functions are presented.

\begin{figure}
\begin{center}
\includegraphics[scale=0.6]{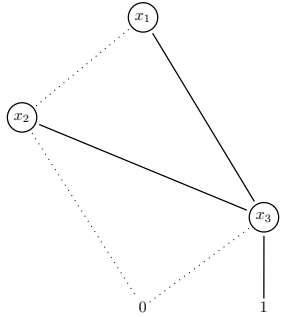}
\end{center}
\caption{The reduced version of the BDD previously shown in Figure~\ref{bdd_unreduced}.}
\label{reduced_bdd}
\end{figure}

The representation which is shown in Figure~\ref{bdd_unreduced} is canonical. However, it is completely inefficient since it takes $O(2^n)$ space to represent a boolean function with $n$ variables in memory. ROBDD (or BDD for short) tries to address this problem by eliminating redundancies in unreduced BDDs. For eliminating redundancies and having a canonical representation, following two constraints should be always satisfied in any ROBDD:\\
1. A ROBDD should be ordered, that is, variables should respect a given total order on any path in a ROBDD. 2. a ROBDD should be reduced which means that there are no two sub-graphs in a ROBDD that are isomorphic and also for any node in a ROBDD its low-child should be different from its high-child.

Note that within every node in a ROBDD, level, a pointer to its low child, and a pointer to its high child are saved. Every node of a ROBDD  which is also a ROBDD can be identified uniquely by a triple ($level$, $low$, $high$). Figure~\ref{reduced_bdd} shows the same boolean function as in Figure~\ref{bdd_unreduced} but in reduced form. We can also see this BDD and its associated boolean function as a set which contains $011$, $101$ and $111$ strings.\\
It is very common to use the term BDD to refer to ROBDD and we follow this practice in the rest of this paper.

\subsection{BuDDy}
\label{buddy_sec}
BuDDy \cite{buddy} is a library for creating and manipulating BDDs. It is written in C and also offers a C++ interface. Since we extended this library, it is useful to know some of its internals which affected our design.

In BuDDy all nodes (BDDs) are stored in an array which is named \verb=bddnodes=. Every slot in this array has four fields namely \verb=level=, \verb=low=, \verb=high= which are used to identify the BDD stored in the slot and the fourth field \verb=hash= which is used to make searching the array more efficient by using hashing.

%\subsection{Reduction in BuDDy}
Every BDD can be uniquely identified by using its \verb=level=, \verb=low= and \verb=high= attributes. In another word, we can associate a triple $(level, low, high)$ with every BDD. At the core of BuDDy is a routine named \verb=bdd_makenode= which is used for allocating BDDs. This routine only creates one entry for every distinct triple in \verb=bddnodes= array and if it is asked to create a triple which is already inserted in \verb=bddnodes=, it simply returns the index of the existing entry. This index represents the BDD in BuDDy. Also, if the triple sent to this routine contains the same value as its \verb=low= and \verb=high=, no new BDD will be allocated and the \verb=low= will be returned.\\
In this way all the BDDs are always reduced and share sub-graphs that are isomorphic. Sub-graphs of any BDD are BDDs themselves and are allocated only once for any distinct triple. This brings some of the advantages of MRBDDs to our implementation. As described in \cite{mrbdd}, BDDs which constitute a MRBDD share isomorphic sub-graphs, but in BuDDy and as a consequence in our implementation any two BDDs share isomorphic sub-graphs.

\subsection{Apply Operation}
BDDs represent boolean functions so one way to manipulate them is through boolean operations like \textbf{or}, \textbf{and}, \textbf{xor}, etc.\\
In general, there is a routine (\verb=bdd_apply= in BuDDy) that takes two BDDs (which represent two boolean functions) and makes a new BDD out of them by applying a boolean operator. For further details see \cite{bdd,buddy} and also Subsection~\ref{intersection_union}.
%In Figure~\ref{bdd_apply}, you see a general routine in pseudo code that takes two BDDs (which represent two boolean functions) and makes a new BDD out of them by applying a boolean operator to them. For further details see \cite{bdd, buddy}.

\subsection{Multi-Terminal Binary Decision Diagrams}
\label{mtbdd_sec}
%~\ref{extend_buddy_sec} ~\ref{mtbdd_fuzzyrelation} ~\ref{extend_buddy_sec}
A Multi-Terminal Binary Decision Diagram (or MTBDD for short) is a data-structure which has all the features of BDD and it allows more than two terminals. In Sections~\ref{extend_buddy_sec},~\ref{mtbdd_fuzzyset_sec} and~\ref{mtbdd_fuzzyrelation}, we explain how we have extended BuDDy to add support for MTBDDs as well as other functionalities which were needed.

\section{Extending BuDDy to support MTBDDs}
\label{extend_buddy_sec}
In BuDDy, \verb=BDD= type is defined as \verb=int=. The integer representative of a BDD can be used to index into \verb=bddnodes= array and retrieve its root. However, terminals are not required to be stored in \verb=bddnodes= array explicitly since integers zero and one are reserved to show them. Integers greater than one are used to show non-terminals. %(BDDs other than terminals zero and one).

We chose not to define any new type to show MTBDD and simply used integer as their representative to comply with existing design. As a result, integers were also used to show terminals other than zero and one. However, the routine \verb=bdd_makenode= can use any slot with index greater than one in \verb=bddnodes= array to store a BDD (a non-terminal) and returns the slot's index as the BDD's representative. Thus using integers greater than one for showing terminals could introduce new complexities in this routine.\\
To overcome this problem, negative integers were used to show terminals other than zero and one (-1 can not be used to show a terminal since it indicates an uninitialized slot in \verb=bddnodes=). In this way, all the existing routines continue to work (except \verb=bdd_apply= in cases that it encounters terminals other than zero and one).

The BuDDy library was extended in a generic way so it can be used in similar scenarios. Major routines which are added to the library are \verb=mtbdd_apply= and \verb=mtbdd_maxmin_compose= (\verb=mmc= for short). The former was added to handle maximum and minimum operators for MTBDDs, and the latter is simply a new functionality which was added to do max-min compositions of two binary fuzzy relations. See Subsection~\ref{max_min_composition} for further detail.

Floating points were not used to show the membership values since imprecision in floating-points was not acceptable for us and we would like to have fully deterministic results. A C struct which has a field of type Integer is used to represent membership values. For example, an instance of this struct with an integer set to 25 shows 0.025 when precision of three digits is used.\\
It may be worth noting that the precision should be known in advance to interpret a membership value. We used three different precisions in our benchmarks. Precision of three digits which can show 1001 different membership values, precision of two digits which can show 101 different membership values and precision of a single digit which can show 11 different membership values (Note that 1 is considered to be a membership value)
\begin{comment}
A C struct which is shown in Figure~\ref{membership_value_struct} is used instead.

\begin{figure} [ht]
\begin{verbatim}
typedef struct {
  unsigned char is_one;
  unsigned short value;
} membership_value;
\end{verbatim}
\caption{The C struct which is used to represent membership values.}
\label{membership_value_struct}
\end{figure}

Considering this data-type whenever \verb=is_one= is 1 we are sure that the membership value is 1 otherwise \verb=value= is the determinant factor. For example, to show membership value 0.255, \verb=is_one= is set to 0 and \verb=value= is set to  255.\\
It may be worth noting that the precision should be known in advance. We have used three different precisions in our benchmarks. Precision of three digits which can show 1001 different membership values, Precision of two digits which can show 101 different membership values and precision of a single digit which can show 11 different membership values (Note that 1 is considered to be membership value). As an example, consider the precision of three digits, in this case, a \verb=membership_value= with 1 as its \verb=value= and 0 as its \verb=is_one= is interpreted as 0.001. The same case is interpreted as 0.01 when precision of two digits is used.
\end{comment}

\section{MTBDDs as Fuzzy Sets}
\label{mtbdd_fuzzyset_sec}
\begin{figure} [ht]
\begin{center}
\includegraphics[scale=0.5]{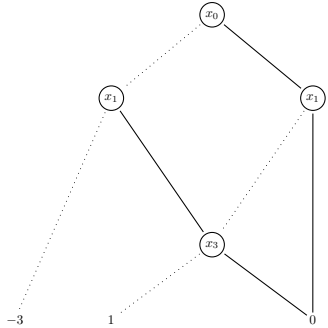}
\end{center}
\caption{A MTBDD representing a fuzzy set.}
\label{mtbdd_fuzzyset}
\end{figure}

In the MTBDD representation of a fuzzy set, there are as many terminals as there are different membership values in the fuzzy set (including zero and one). Different paths (including those which are reduced or are not represented explicitly) show different members of the fuzzy set. The terminal each path ends at, shows its membership value. In Figure~\ref{mtbdd_fuzzyset}, membership value 0.3 is represented by -3 so the MTBDD shows the fuzzy set \{0.3/0000, 0.3/0001,\\ 0.3/0010, 0.3/0011, 1/0100, 1/0110, 1/1000, 1/1010\}.\\Strings $0000, 0001, 0010, ...$ correspond to numbers $0, 1, 2, ...$ respectively.

\subsection{Intersection and Union operations}
\label{intersection_union}
Maximum and minimum operators were used as fuzzy set intersection and fuzzy set union respectively \cite{zadeh}. The general apply routine which was mentioned in Section~\ref{background} takes two BDDs as its operands and another parameter as its operator, then, it applies the operator to the operands. The apply routine descends through its operands (from root to terminals) and makes necessary nodes and links in the resulted BDD along the way. When the routine encounters terminals (in the generic case, one of the terminal is from the first BDD operand and the other one is from the second BDD operand), it chooses a function from a table of functions, based on the operator, and applies the selected function to the terminals. The resulted terminal is associated with the same path that the two terminals from the first and second operand is associated with.

In our implementation this table of functions was extended to include functions for maximum and minimum operators. The functions which were added can handle terminals with values one and zero as well as terminals with other values. A slightly modified apply routine (\verb=mtbdd_apply=) which handles MTBDDs and maximum/minimum operators was added to the BuDDy library.

\section{MTBDDs as Binary Fuzzy Relations}
\label{mtbdd_fuzzyrelation}
Any binary fuzzy relation has two domains, two disjoint sets of BDD variables are used. Every set of BDD variables is mapped to one of the domains. For example, if a domain has eight objects, three variables would be needed to show all its members (i.e. $2^3 = 8$).

\begin{figure}
\begin{center}
\includegraphics[scale=0.8]{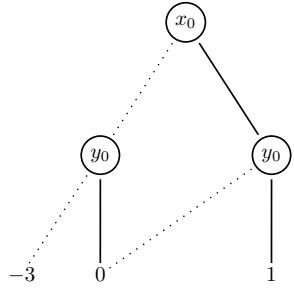}
\end{center}
\caption{A MTBDD which represents the binary fuzzy relation \{$1/(1,1), 0.3/(0,0)$\}}
\label{small_relation}
\end{figure}

In Figure~\ref{small_relation}, there are two domains and each one has two objects so all objects can be encoded by using one variable for each domain. Variable $x_0$ is used to encode the objects in the first domain and variable $y_0$ is used to encode objects in the second domain. Suppose that non-terminal $-3$ is mapped to $0.3$. In this way ($0, 1$) and ($1, 0$) are associated with $0$ membership value. ($1, 1$) is associated with $1$ membership value and ($0, 0$) is associated with $0.3$ membership value. \\
You can also see it as the 2-dimensional array:\\
\begin{center}
$
\begin{pmatrix}
0.3 & 0\\
0 & 1
\end{pmatrix}
$
\end{center}

\begin{figure}
\begin{center}
\includegraphics[scale=0.6]{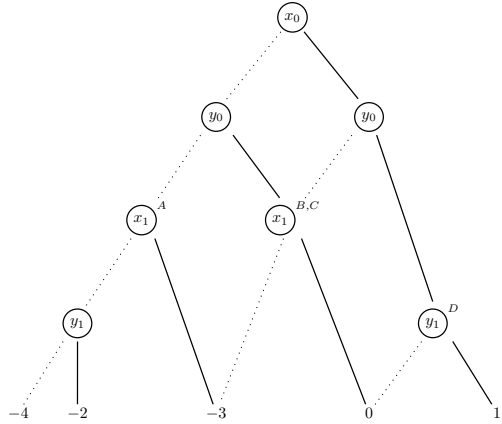}
\end{center}
\caption{A binary fuzzy relation represented as a MTBDD. Nodes A, B, C and D show four partitions of this MTBDD. Both B and C correspond to the same node.}
\label{big_relation}
\end{figure}

Since we implemented an algorithm similar to 4-way block-multiplication to compute the max-min composition of two binary fuzzy relations which are represented as MTBDDs, it is desirable to partition each relation into four blocks and access each one in constant time. In order to make this possible, interleaved variable ordering was used \cite{mtbdd}. This means that if variables $x_i$ constitute domain $x$ and variables $y_i$ constitute domain $y$, the ordering of variables would be $x_0,y_0,x_1,y_1,x_2,y_2,...$. See Figure~\ref{big_relation} for an example.\\
Binary fuzzy relations can be seen as square matrices of size $n \times n$. In a binary fuzzy relation $A$ that $n \neq 2^k $, an identity matrix with smallest possible size is attached to $A$ to meet this requirement:\\
\begin{center}
$
\begin{pmatrix}
A & 0 \\
0 & I
\end{pmatrix}
$
\end{center}
This technique is also used in \cite{mtbdd}. This is working for max-min composition since minimum of zero and any other membership value is zero (similar to matrix multiplication and multiplication of zero by other element).

\subsection{Max-min composition}
\label{max_min_composition}
In general max-min composition of two binary fuzzy relations $R_1$ ($D \times D$) and $R_2$ ($D \times D$) is defined as follows:\\
$R_3(a,b) = max_{c \in D} min(R_1(a, c), R_2(c, b))$ 

The max-min composition procedure is similar to block matrix multiplication. A binary fuzzy relation can be viewed as a 2-dimensional matrix. This matrix is partitioned into four sub-matrices (blocks) in the procedure(Figure~\ref{big_relation}). The partitioning of the relation into four blocks is done in constant time since each partition of a MTBDD is at most two hops away from its root.\\
max-min composition was implemented as a recursive procedure which is shown in Figure~\ref{mmc}. During the recursion, at each call, parameters of the call (MTBDD \verb=a= and MTBDD \verb=b=) should be interpreted based on the depth of the recursion which is passed as the third parameter (\verb=call_level=). This is because, the partitioning does not create four new MTBDDs but returns four sub-graphs of the original MTBDD as its partitions so a hypothetical level (\verb=root_level=) is assumed. The mentioned hypothetical level indicates the smallest possible level of the resulted partitions (Figure~\ref{mmc}). For example, consider the MTBDD shown in Figure~\ref{big_relation}, four partitions would be created after partitioning namely A, B, C and D. The hypothetical root level for these partitions is two which corresponds to the variable $x_1$. This technique (introducing and using a hypothetical root level) avoids creation of new matrices (MTBDDs) and makes the max-min composition procedure more efficient.\\
In order to compute the max-min composition of MTBDDs \verb=a= and \verb=b=, we have to call \verb=mmc(a, b, 0)=. If \verb=a= and \verb=b= do not correspond to square matrices, they should be extended to binary fuzzy relations of size $2^k \times 2^k$ as described before.

\begin{figure}
\begin{algorithmic}[1]
\Procedure{mmc}{$a,b,callLevel$}

\If{both $a$ and $b$ are terminals} 
\State \textbf{return} $min(a, b)$
\EndIf

\If{($r \gets$mmc-cache[$a, b, callLevel$]) $\neq$ NULL}
\State  \textbf{return} $r$
\EndIf

\If{($r \gets$mmc-cache[$b, a, callLevel$]) $\neq$ NULL}
\State  \textbf{return} $r$
\EndIf
\\
\State  $rootLevel \gets callLevel * 2$
\\
\State  partition a into $sa[0], sa[1], sa[2]$ and $sa[3]$ based on 
  $rootLevel$
\\
\State  partition b into $sb[0], sb[1], sb[2]$ and $sb[3]$ based on 
  $rootLevel$
\\
\State  $t1 \gets$ MMC($sa[0], sb[0], callLevel+1$)
\State  $t2 \gets $MMC($sa[1], sb[2], callLevel+1$)
\State  $l \gets$ MTBDD-APPLY($t1, t2, mtbddopFuzzymax$)
\State  $t1 \gets$ MMC($sa[0], sb[1], callLevel+1$)
\State  $t2 \gets$ MMC($sa[1], sb[3], callLevel+1$)
\State  $h \gets$ MTBDD-APPLY($t1, t2, mtbddopFuzzymax$)
\State  $L \gets$ BDD-MAKENODE($rootLevel+1, l, h$)
\\
\State  $t1 \gets$ MMC($sa[2], sb[0], callLevel+1$)
\State  $t2 \gets$ MMC($sa[3], sb[2], callLevel+1$)
\State  $l \gets$ MTBDD-APPLY($t1, t2, mtbddopFuzzymax$)
\State  $t1 \gets$ MMC($sa[2], sb[1], callLevel+1$)
\State  $t2 \gets$ MMC($sa[3], sb[3], callLevel+1$)
\State  $h \gets$ MTBDD-APPLY($t1, t2, mtbddopFuzzymax$)
\State  $H \gets$ BDD-MAKENODE($rootLevel+1, l, h$)
\\
\State  $r \gets$ BDD-MAKENODE($rootLevel, L, H$)
  
\State  mmc-cache[$a, b, callLevel$] $\gets r$
\State  \textbf{return} $r$

\EndProcedure
\end{algorithmic}
%
\begin{comment}
\begin{verbatim}
mmc(a, b, call_level) {
  if both a and b are terminals return min(a, b);
  
  if ((r=mmc_cache(a, b, call_level))!=NULL))
    return r;
  if ((r=mmc_cache(b, a, call_level))!=NULL))
    return r;
  root_level = call_level * 2;
  
  partition a into sa[0], sa[1], sa[2] and sa[3] based on 
  root_level;

  partition b into sb[0], sb[1], sb[2] and sb[3] based on 
  root_level;

  t1 = mmc(sa[0], sb[0], call_level+1);
  t2 = mmc(sa[1], sb[2], call_level+1);
  l = mtbdd_apply(t1, t2, mtbddop_fuzzymax);
  t1 = mmc(sa[0], sb[1], call_level+1);
  t2 = mmc(sa[1], sb[3], call_level+1);
  h = mtbdd_apply(t1, t2, mtbddop_fuzzymax);
  L = bdd_makenode(root_level+1, l, h);

  t1 = mmc(sa[2], sb[0], call_level+1);
  t2 = mmc(sa[3], sb[2], call_level+1);
  l = mtbdd_apply(t1, t2, mtbddop_fuzzymax);
  t1 = mmc(sa[2], sb[1], call_level+1);
  t2 = mmc(sa[3], sb[3], call_level+1);
  h = mtbdd_apply(t1, t2, mtbddop_fuzzymax);
  H = bdd_makenode(root_level+1, l, h);

  r = bdd_makenode(root_level, L, H);
  
  mmc_cache(a, b, call_level) = r;
  return r;
}
\end{verbatim}
\end{comment}
%
\caption{Max-min composition (mmc) routine in pseudo code. $a$ and $b$ are MTBDDs and $callLevel$ indicates the depth of the recursion.}
\label{mmc}
\end{figure}

\section{Evaluation}
\label{evaluation}

To evaluate our representation, we extended the BuDDy library to represent and manipulate binary fuzzy relations by using MTBDDs. Also, we implemented binary fuzzy relations based on two dimensional arrays. The array implementation was used as a baseline (base implementation).\\
Images in our input set are obtained from UIUC image database \cite{uiuc_bench} and The Berkeley Segmentation Dataset and Benchmark \cite{berk_bench}.

In our first experiment, max-min composition procedure was evaluated. In the second experiment, the fuzzy-connectedness problem is solved for different images in the input set.\\
Results of these experiments and further details come next.

\subsection{Evaluation results and further details}

\begin{figure*}
%\begin{center}
\centering \includegraphics[angle=270,scale=0.40]{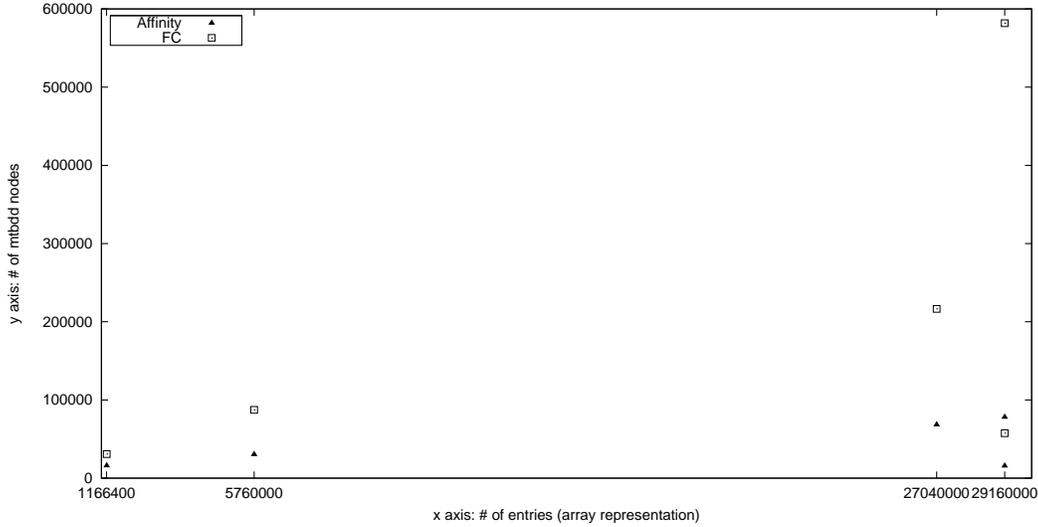}
%\end{center}

\caption{Compares number of entries in the array representation of relations with number of nodes in their equivalent MTBDD representations.}
\label{nodes_graph}

\end{figure*}

Tables~\ref{mmc_times} and~\ref{mmc_nodes} give the results of our first experiment. Results for the second experiment (fuzzy-connectedness problem) are given in tables~\ref{conn_times},~\ref{conn_nodes} and ~\ref{conn_nodes_kb}. Figure~\ref{nodes_graph} illustrates results from both experiment. Input relations for our experiments, are affinity relations, which are created from various images. Affinity relation is a symmetric and reflexive fuzzy relation which assigns a membership value to a pair of pixels based on their local properties \cite{conn}. We initialized this relation only for pairs of pixels which are neighbor in a given image and membership value for any other pair of pixels in the image is set to zero. This leads to sparsity of affinity relations.\\
The following similarity measure which was used in \cite{graph_seg} has been employed to initialize affinity relations. $\delta$ is the largest \textit{diff} (\textit{diff} is computed for every pair of pixels) and $c_r$, $c_g$, $c_b$, $d_r$, $d_g$ and $d_b$ show color intensities associated with $c$ and $d$ pixels respectively:\\ 
$simil(c,d) = 1 - \sqrt{diff} / \delta$ where $diff=(c_r-d_r)^2 + (c_g-d_g)^2 + (c_b-d_b)^2$.\\
Images  used for creating affinity relations are shown in Table~\ref{images}. The first one is from UIUC image database \cite{uiuc_bench} and the next three images are obtained from The Berkeley Segmentation Dataset and Benchmark \cite{berk_bench}. The last image is a synthetic image from reference \cite{graph_seg}.

All experiments are run on a machine with 2.8 GHz Intel CPU and 4 GB of RAM running Fedora 14.

\begin{table}
\begin{center}
\begin{footnotesize}

\caption{Images which are used for creating affinity relations. }
\label{images}

%\begin{comment}
\begin{tabular} {| c | c | c |}
\hline
image & image name & size  \\ \hline
\includegraphics[scale=.6]{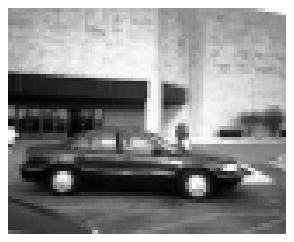} & $image_0$ & 80x65 \\ \hline
\includegraphics[scale=.6]{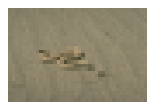} & $image_1$ & 40x27 \\ \hline
\includegraphics[scale=.6]{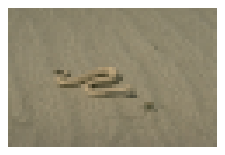} & $image_2$ & 60x40 \\ \hline
\includegraphics[scale=.6]{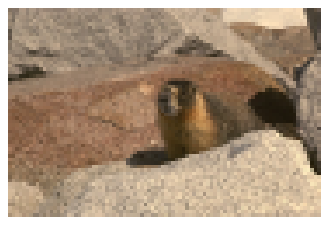} & $image_3$ & 90x60 \\ \hline
\includegraphics[scale=.6]{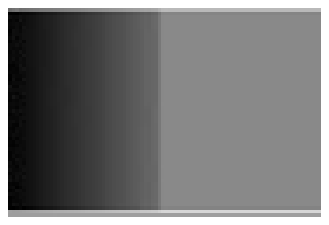} & $image_4$ & 90x60 \\ \hline
\end{tabular}
%\end{comment}
 
\end{footnotesize}
\end{center}

\end{table}

In the first experiment, for every binary fuzzy relation \verb=r=, \verb=mmc(r, r)= is computed using both our data-structure (based on MTBDD) and the base implementation (based on two dimensional arrays). Running times of both base implementation and MTBDD based implementation are shown in Table~\ref{mmc_times}. Column base, shows the running time of base implementation and the other two columns under mtbdd, show running-times for two different precisions of mtbdd implementation. (1) and (2) indicate precision of one and two digits respectively.\\
As you can see MTBDD based implementation performs significantly better in all cases, specially when image's pixels are more homogeneous or the image is larger (it is faster by a factor ranging from 164 to 1328). Two different versions of the MTBDD  based implementation was run, one with precision of one digit and the other with precision of two digits. Number of terminals in the former is limited to 11 ($0.0, 0.1, 0.2, ..., 1.0$) and number of terminals in the latter is limited to 101 ($0.00, 0.01, 0.02, ..., 1.00$). Note that terminals are not pre-allocated in our implementation but are simply represented by integers ($0,1$ and $-2, -3, -4, ...$) in \verb=bddnodes= table.\\
Numbers of nodes and terminals in resulted MTBDDs are shown in Table~\ref{mmc_nodes}. A comparison of number of nodes in resulted MTBDDs with number of entries in their equivalent relations in base implementation shows a huge difference. This leads to an extremely improved memory consumption in MTBDD implementation. It also explains why the MTBDD version is faster, as absence of redundancies in MTBDD representation leads to elimination of unnecessary computations. 
Considering that affinity relations (and to some degree resulted relations produced by calling the \verb=mmc= procedure) are sparse, Another experiment (solving fuzzy-connectedness problem) which results in full binary fuzzy relations was conducted. The experiment is described next.

\begin{table}
\begin{center}
\begin{footnotesize}

\caption{Running times of mmc routine for both base and MTBDD based implementations.}
\label{mmc_times}

\begin{tabular} {|c|c| cc |}
\hline
image & base & \multicolumn{2}{c}{mtbdd} \vline \\

      &      & (1)     &     (2) \\ \hline 
$image_0$ & 3152 & 2.42 &  3.56 \\ \hline
$image_1$ & 27.93 & 0.11 & 0.17 \\ \hline
$image_2$ & 307.44 & 0.39 & 0.73 \\ \hline
$image_3$ & 3534 & 2.66 & 4.14 \\ \hline
$image_4$ & same as $image_3$ & 1.81 & 2.22\\ \hline
      
\end{tabular}
\end{footnotesize}
\end{center}

\end{table}

%%%%%%%%%%%%%%%%%%%%%%%%%%%%%%%%%%%%%%%%%%%%%%
\begin{table}
\begin{footnotesize}

\caption{Number of entries in resulted relations from mmc operation and number of nodes allocated in MTBDD representation of these relations}
\label{mmc_nodes}

\begin{center}
\includegraphics[scale=0.8]{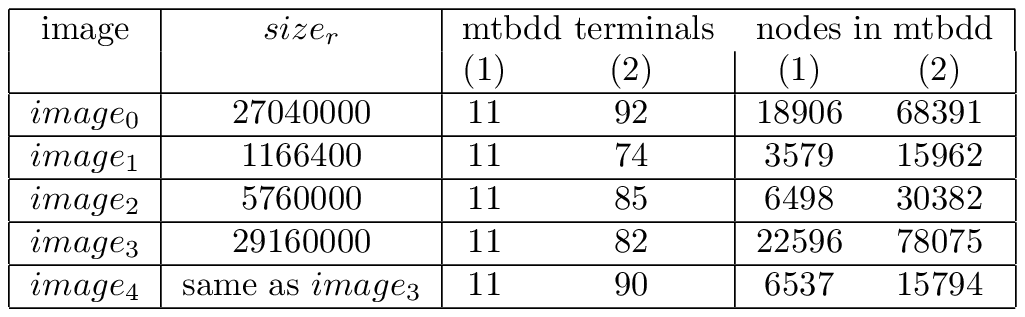}
\end{center}

\end{footnotesize}
\end{table}
%%%%%%%%%%%%%%%%%%%%%%%%%%%%%%%%%%%%%%%%%%%%%%%%%

%%%%%%%%%%%%%%%%%%%%%%%%%%%%%%%%%%%%%%%%%%%%%%%%

\begin{table}

\begin{center}
\begin{footnotesize}

\caption{Running times of fuzzy connectedness problem solver for both base and MTBDD based implementations.}
\label{conn_times}

\begin{tabular} {|c|c| cc |}
\hline
image & base & \multicolumn{2}{c}{mtbdd} \vline \\
      &   & (1)  & (2) \\ \hline
$image_0$ & 3574 & 134.61 & 2236 \\ \hline
$image_1$ & 32.01 & 1.78 & 41.90\\ \hline
$image_2$ & 350.72 & 13.88 & 285.48 \\ \hline
$image_3$ & 4000 & 214.64 & 1898.21 \\ \hline
$image_4$ & same as $image_3$ & 148.95 & 533.52 \\ \hline

\end{tabular}

\end{footnotesize}
\end{center}
\end{table}
%%%%%%%%%%%%%%%%%%%%%%%%%%%%%%%%%%%%%%%%%%

\begin{table}

\begin{footnotesize}
\caption{Number of entries in FC relations and number of nodes which are allocated
to represent the relation in its MTBDD representation}
\label{conn_nodes}

\begin{center}

\includegraphics[scale=0.8]{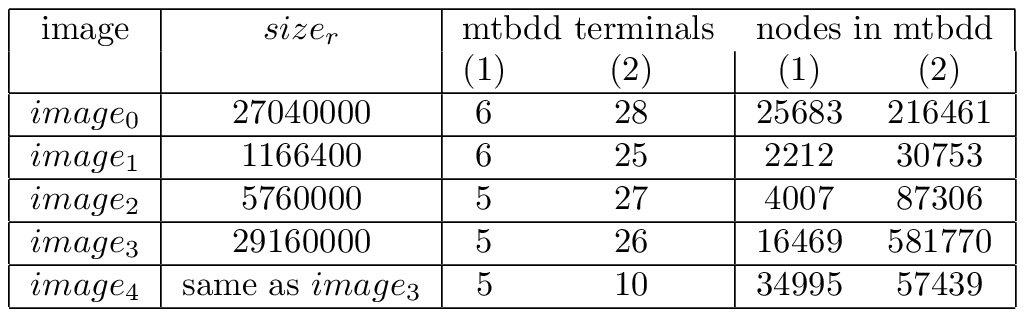}
\end{center}

\end{footnotesize}
\end{table}

In the rest of this Section, problem of fuzzy-connectedness is investigated (our second experiment). Input to this algorithm is an affinity relation which is extracted from an image, and final output is a relation that assigns a membership value to every pair of pixels in the image. The output can be used to create different clusters of pixels \cite{conn}.\\
The final goal of fuzzy connectedness is to calculate FC relation which is a reflexive, symmetric and transitive relation. It is basically max-min transitive closure of the initial affinity relation. In contrast to affinity relation, FC relation is a full binary fuzzy relation. It assigns a membership value to every pair ($c, d$). This value is the maximum strength of all possible paths from $c$ to $d$. The strength of a path is the smallest membership value along the path. FC relations are obtained by computing max-min transitive closure of affinity relations.

We implemented two different versions to compute the transitive closure. Our base implementation used two dimensional arrays to represent binary fuzzy relations and, it employed Floyd-warshall algorithm as shown in Figure~\ref{floyd}. $n$ is the number of pixels in the image. $c$ stores the affinity relation initially and represents fuzzy connectedness (FC) relation at the end.\\
Our second implementation used MTBDDs to represent binary fuzzy relations and, it computed FC relation by using Repeated Squaring algorithm as shown in Figure~\ref{repeated_mmc}. Affinity relation is the input to this algorithm and at the end, $res$ would be a MTBDD that represent the fuzzy connectedness relation (FC).\\
Because of the MTBDD special structure we could not use the Floyd-warshall algorithm in conjunction with this data-structure efficiently.

Table~\ref{conn_times} shows running-times of our base and MTBDD based programs which compute max-min transitive closure of affinity relations obtained from our test images. Three versions are shown in the table. Column base shows the base implementation and the other two columns under mtbdd show the MTBDD based implementation with one and two digits of precision. The MTBDD based implementation is significantly faster than the base implementation when precision of one digit is used (it is 18 -- 27$\times$ faster). Compared to base implementation, using 2-digits precision improved running time in all cases, but one, which was our smallest image ($40 \times  27$). In this particular case all running times were under one minute. In other cases MTBDD based implementation with two digits precision is faster by a factor ranging from 2 to 7. 

When images are larger and their pixels are more homogeneous, MTBDD based implementation becomes faster relative to the base implementation. Note that running-time in base implementation only depends on size of input image (i.e. number of pixels). In contrast, MTBDD based implementation's running-time depend both on size of image and values stored in every pixel of the image. For example, running-times for $image_3$ and $image_4$ are the same when base implementation is used but it takes less time for MTBDD based implementation to compute the transitive closure when it takes $image_4$ as input. 

As described in Section~\ref{mtbdd_fuzzyrelation}, different paths in MTBDD representation of a relation show different pairs in the relation. More commonalities among paths lead to more reductions and, a more compact MTBDD representation of the relation. Computations on a smaller MTBDD takes less time. Two different images even with the same size results in different MTBDDs. An image which results in a MTBDD with more commonalities in its paths occupies less memory and results in fewer computations in the MTBDD based solver (Sparsity in input relation and also images with homogeneous pixels leads to more compact MTBDDs).

In Table~\ref{conn_nodes}, number of entries in FC relations, number of terminals and number of nodes in MTBDD representation of these relations are shown. Required computations are done as shown in Figures~\ref{floyd} and~\ref{repeated_mmc} for base and MTBDD based implementations respectively.\\ 
Number of nodes in MTBDD based implementation is extremely lower than number of entries in base implementation which leads to a far improved memory consumption when MTBDDs are used to represent binary fuzzy relations. Note that, resulted relations (FC) are full in contrast to the previous experiment. Every array's entry in base implementation is three bytes (two bytes for a short integer and one byte for a flag) and the size of every bddnode is 20 bytes \cite{buddy}. In Table~\ref{conn_nodes_kb} size of array and MTBDD based representation of FC relations are shown (in KB). The column $size_r$ indicates number of array's entry in the relation, column size(array) shows the amount of memory allocated for representing arrays in the base implementation and, the other two columns indicate the amount of memory allocated for representing MTBDDs in the MTBDD based implementation (precision of one and two digits). Considering this table, MTBDD based representation takes 37.9 -- 265.5$\times$ less memory than the array representation depending on the input image.
%There is a direct relation between number of nodes in MTBDD representation of relations and time it takes to compute them.
\begin{table}
\begin{footnotesize}
\caption{Size of array and MTBDD representation in KB}
\label{conn_nodes_kb}

\begin{center}
\begin{tabular} {| c | c | c | cc | }
\hline
image  &  $size_r$  & size(array) & \multicolumn{2}{c}{size(mtbdd)} \vline \\
       &            &             &  (1)  &  (2)  \\ \hline

$image_0$   &   27040000   & 81120 &  513 & 4329 \\ \hline
$image_1$   &   1166400    & 3499 &  44 & 615 \\ \hline
$image_2$   &   5760000    & 17280 &  80 & 1746 \\ \hline
$image_3$   &   29160000   & 87480 &  329 & 11635 \\ \hline
$image_4$   &   29160000   & 87480 &  699 & 1148 \\ \hline
\end{tabular} 
\end{center}

\end{footnotesize}
\end{table}

It is also noteworthy that shape and number of nodes in BDDs (and MTBDDs as well) also depend on variable ordering beside data they are representing since different variable ordering leads to different paths with different degree of sharing. However, a fixed variable ordering is used in our implementation.

Number of terminals are also shown in Table~\ref{conn_nodes}. This gives the number of distinct (hard) clusterings that can be obtained from the resulted relation. When number of terminals is limited to 11 (1-digit precision), it is five or six and in the other case, when number of terminals is limited to 101 (2-digit precision), it is usually around 25.\\
Number of terminals in Table~\ref{conn_nodes} is less than number of terminals in Table~\ref{mmc_nodes}. Considering definition of FC relation, when there are multiple paths between two pixels, the largest strength among them is chosen as the membership value for the pair of pixels in the FC relation so some membership values might be eliminated through this process. Note that there are as many terminals as distinct membership values in the FC relation.

\begin{figure}[ht]

\begin{algorithmic}[1]
\For{$k \gets 1$ \textbf{to} $n$}
  \For{$i \gets 1$ \textbf{to} $n$}
    \For{$j \gets 1$ \textbf{to} $n$}
      \State $c[i, j] \gets max(c[i, j], min(c[i, k], c[k, j]))$
    \EndFor
   \EndFor
\EndFor

\end{algorithmic}
\caption{Using Floyd-warshall algorithm to compute max-min transitive closure of affinity relation.}
\label{floyd}
\end{figure}

\begin{figure}[ht]  
\begin{algorithmic}[1]
\State $res \gets affinity$
\Repeat 
  \State $old \gets res$
  \State $res \gets$ MMC($res, res$) \Comment MMC(res, res, 0)
\Until{$res = old$}
\end{algorithmic}
%
\begin{comment}
\begin{verbatim}
/* At the end, res would be a MTBDD that represent the fuzzy connectedness
   relation.
*/
res = affinity;
do 
  old = res;
  res = mmc(res, res); /* i.e. mmc(res, res, 0) */
while (res != old)
\end{verbatim}
\end{comment}
%
\caption{Using MTBDDs and Repeated Squaring algorithm to compute max-min transitive closure of affinity relation.}
\label{repeated_mmc}
\end{figure}

The graph in Figure~\ref{nodes_graph} compares number of entries in the array representation of relations with the number of nodes in their equivalent MTBDD representations. In this Figure, axis $x$ is associated with the number of entries in the base implementation (array representation) and axis $y$ is associated with the number of nodes in the MTBDD representation. Both affinity (which are sparse) and FC (which are full) relations are considered and number of nodes for MTBDDs of two digits are shown.

MTBDD size increases as images get larger (with the exception of $image_0$ and $image_4$) but due to compactness of MTBDDs, they scale well. Number of MTBDD nodes does not only depend on image size but the value stored in every pixel, and the pattern that pixels are scattered (i.e. order of pixels) too. As explained earlier in this Section, these parameters lead to different MTBDDs with different sizes and the ones which have paths with more commonalities are represented more compactly. For every point in the first three points on the $x$ axis, there are two points on $y$ axis, one for the FC relation and another for the affinity relation. However, for the last point on the $x$ axis, there are four points on the $y$ axis. This is because, there are two images of the same size ($image_3$ and $image_4$), and both images correspond to the same point on $x$ axis. Numbers 15794 (Affinity) and 57439 (FC) indicate number of MTBDD nodes for $image_4$ and Numbers 78075 (Affinity) and 581770 (FC) indicate number of MTBDD nodes for $image_3$.

Although, $image_4$ is larger than $image_0$, number of MTBDD nodes representing $image_4$ relations are less than number of MTBDD nodes representing $image_0$ relations. Considering these two images, number of MTBDD nodes for FC relation is decreased from 216461 ($image_0$) to 57439 ($image_4$), and number of MTBDD nodes for affinity relation is also decreased (from 68391 ($image_0$) to 15794 ($image_4$)). This perfectly shows that size of image is not the only factor in determining the number of MTBDD nodes (value stored in every pixel, order of pixels and BDD variables order are three other important factors).
%Different patterns lead to different MTBDD encodings that are reduced in different ways and have different sizes as a result. For instance both $image_3$ and $image_4$ are of the same size but more nodes are allocated to represent $image_3$'s affinity and FC relation. As another example, we can see that MTBDD representation of affinity relations, have more potential for reduction and are smaller in size compared to MTBDD representation of FC relations since the formers are sparse and the latters are full.
%todo: say that mmc is faster because mtbdds are smaller

\section*{Conclusion}
\label{conclusion}
%In this work, we have designed and implemented a new representation for fuzzy sets and relations which is based on MTBDDs and showed its effectiveness by applying it on fuzzy connectedness and image segmentation problem.
%We also would like to explore the idea of applying similar methods to fuzzy relational databases \cite{frd}.

In this work, we designed and implemented a MTBDD based data-structure to represent fuzzy sets and relations. Also, the BuDDy library was extended to support MTBDDs, and it was employed to implement our idea.\\
Effectiveness of our methods have been shown for both sparse and full binary fuzzy relations. The former was performed by doing a single max-min composition operation on different sparse relations and the latter was performed by solving fuzzy connectedness problem for different images. The solution to fuzzy connectedness problem is a full relation. Promising results have been obtained. In particular, considering the fuzzy connectedness problem and compared to the base implementation, when MTBDD based implementation was used, the running-time was improved by a factor ranging from 2 to 27 and, 37.9 -- 265.5$\times$ less memory was required to represent the final results.

In the future we would like to apply our data-structure to other problems in fuzzy systems which involve manipulating binary fuzzy relations and fuzzy sets, specially, problems with very large sets of fuzzy data such as the use of fuzzy sets in data mining, approximate reasoning and information retrieval based on fuzzy logic \cite{fdm,fim,supervised_fuzzy_classification,sabio}.
%Also, it is in our interest to explore the image segmentation problem in special cases like image segmentation of noisy images \cite{cmeans_noisy, gibbs_noisy}. These cases might need other similarity measures and methods than the ones presented in this work. This would  affect the MTBDD representation of relations and thus the overall performance.\\
%Also, it is in our interest to explore construction and representation of multi-relationship fuzzy concept and fuzzy ontology networks \cite{concept_networks,fuzzy_ontology} with MTBDDs.
\\

\bibliographystyle{unsrt}
\bibliography{mtbdd-paper}

\end{document}